\documentclass[review]{elsarticle}
\usepackage{lineno,hyperref}
\modulolinenumbers[5]

\journal{Journal of \LaTeX\ Templates}

\biboptions{authoryear}

\usepackage{subfigure}                                                                                 
\usepackage{graphicx}                                                                                  
\usepackage{times}                                                                                     
\usepackage{tabularx}                                                                                  
\usepackage{natbib}                                                                                    
\usepackage{url}                                                                                       
\usepackage{xcolor}                                                                                    
\usepackage{caption}                                                                                   
\usepackage{amsmath}                                                                                   
\usepackage{float}                                                                                     
\usepackage{multirow}                                                                                  
\usepackage{morefloats}                                                                                
\usepackage{amssymb}                                                                                   

\newcommand{\Msun}{\mathrm{M}_{\odot}}

\newcommand{\cm}{\mathrm{cm}}

\newcommand{\MeV}{\mathrm{MeV}}
\newcommand{\GeV}{\mathrm{GeV}}

\newcommand{\TeV}{\mathrm{TeV}}

\definecolor{myred}{rgb}{1,0,0} 
\definecolor{myblue}{rgb}{0,0,1}
\definecolor{mygreen}{rgb}{0,1,0}

\def\stacksymbols #1#2#3#4{\def\theguybelow{#2}
        \def\verticalposition{\lower#3pt}
        \def\spacingwithinsymbol{\baselineskip0pt\lineskip#4pt}
        \mathrel{\mathpalette\intermediary#1}}
\def\intermediary #1#2{\verticalposition\vbox{\spacingwithinsymbol
        \everycr={}\tabskip0pt
        \halign{$\mathsurround0pt#1\hfil##\hfil$\crcr#2\crcr
                \theguybelow\crcr}}}

\begin{document}

\begin{frontmatter}

\title{On the Challenges of Cosmic-Ray Proton Shock Acceleration in the Intracluster Medium}

\author{Denis Wittor}
\address{Hamburger Sternwarte, Gojenbergsweg 112, D-21029 Hamburg, Germany}

\begin{abstract}
 Galaxy clusters host the largest particle accelerators in the Universe: Shock waves in the intracluster medium (ICM), a hot and ionised plasma, that accelerate particles to high energies. Radio observations pick up synchrotron emission in the ICM, proving the existence of accelerated cosmic-ray electrons. However, a sign of cosmic-ray protons, in form of $\gamma$-rays. remains undetected. This is know as the \textit{missing $\gamma$-ray problem} and it directly challenges the shock acceleration mechanism at work in the ICM. 
 
 Over the last decade, theoretical and numerical studies focused on improving our knowledge on the \textit{microphysics} that govern the shock acceleration process in the ICM. These new models are able to predict a $\gamma$-ray signal, produced by shock accelerated cosmic-ray protons, below the detection limits set modern $\gamma$-ray observatories. In this review, we summarise the latest advances in solving the missing $\gamma$-ray problem.
\end{abstract}


\end{frontmatter}


\section{Introduction}

\begin{figure}
 \includegraphics[width = \textwidth]{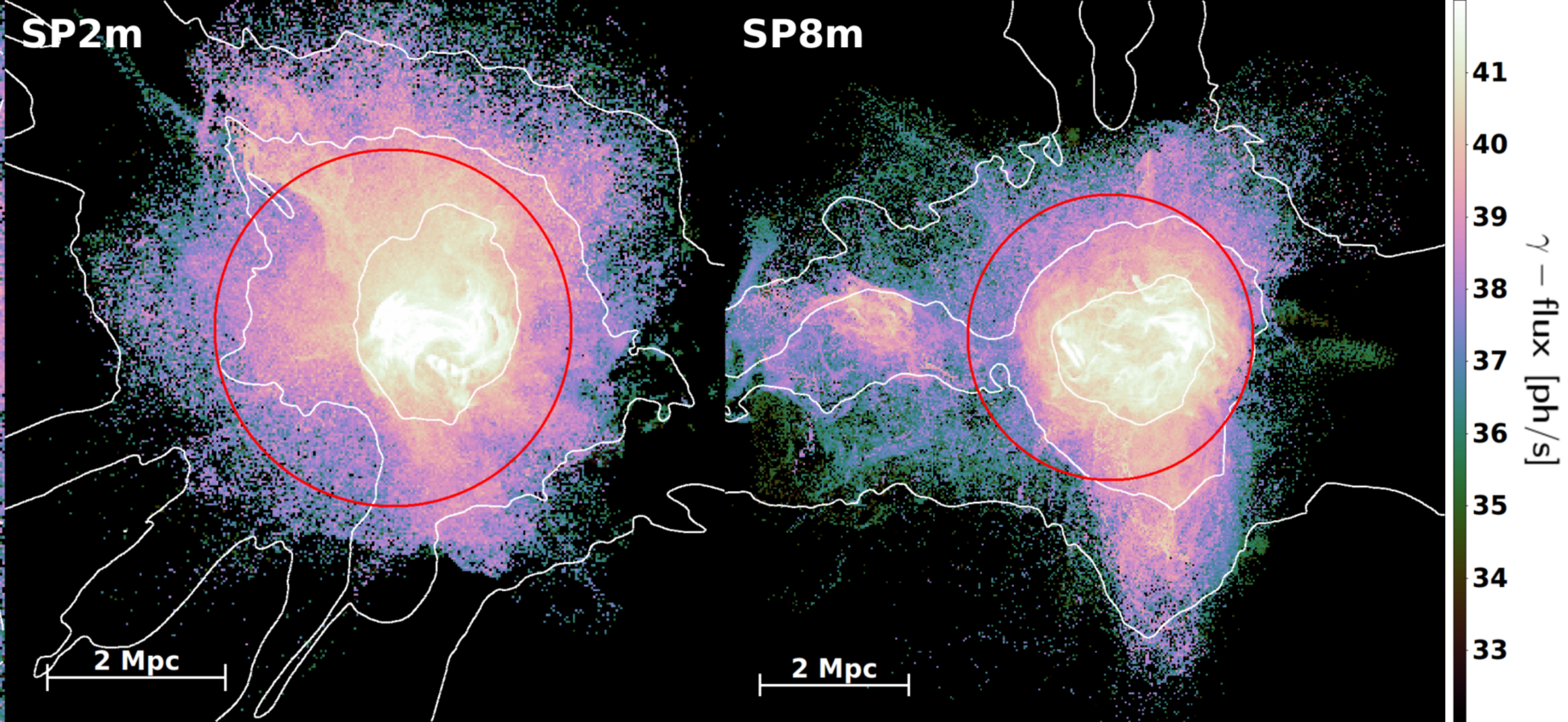}
 \caption{Results from \cite{wittor2020gammas}: the two panels show the expected $\gamma$-ray emission (color) of two simulated galaxy clusters at redshift $z = 0$. The underlying cosmic-ray model is the most restricted one, i.e. only supercritical and quasi-parallel shocks are allowed to accelerate cosmic-ray protons (see Sec. \ref{sec::shocksandprotons}). The white contours five the baryonic density of the ICM at $\left[ 10^{-27.25}, \ 10^{-28}, \ 10^{-28.75} \ \& \ 10^{-29.5}  \right] \ \mathrm{g}/\cm^3$. The red circle marks the $r_{200}$ of each cluster, i.e. the radius that encloses 200 times the critical density.}
 \label{fig::maps}
\end{figure}

Shock waves in the intracluster medium (ICM), the hot and ionised high $\beta$\footnote{Here, $\beta$ is the plasma $\beta$ that gives the ratio between gas pressure and magnetic pressure.} plasma inside galaxy clusters, belong to the largest sites of cosmic-ray acceleration. Galaxy clusters are gravitationally bound systems with masses of $\sim 10^{15} \Msun$. However, only a small portion of their mass, $\lesssim \mathrm{few} \ \%$, consists of galaxies and the main mass contributor is dark matter, $\sim 85 \ \%$. The remaining mass, $\sim 10-15 \ \%$, resides in ICM. During the processes of hierarchical structure formation, galaxy clusters form through the accretion of matter and the merging with other clusters \citep[e.g.][]{2015SSRv..188...93P}. These highly energetic processes drive both shock waves and turbulence in the ICM, giving rise to large acceleration sites for cosmic-rays. Radio observations of diffuse synchrotron emission proof the acceleration of cosmic-ray electrons in the ICM as well as the existence of large-scale magnetic fields \citep[e.g.][]{2008SSRv..134...93F,2012A&ARv..20...54F,vanweeren2019review}. Following the definition by \citep{vanweeren2019review}, diffuse radio sources in the ICM are classified into three groups: 1) giant radio halos and mini-haloes, 2) radio relics (or radio shocks) and 3) revived AGN fossil plasma sources. While the origin of the latter sources is still debated, it is commonly assumed that halos and relics are produced by cosmic-ray electrons that have been (re-)accelerated by ICM turbulence and shocks, respectively. On the other hand, any signatures of cosmic-ray protons in the ICM are still waiting for their detection. In the case of radio halos, the non-detection of cosmic-ray protons favors the turbulent re-acceleration model \citep[e.g.][and references therein]{2001MNRAS.320..365B,2001ApJ...557..560P,donn10,2013MNRAS.429.3564D,2014MNRAS.443.3564D,2017MNRAS.465.4800P} over the hadronic (or secondary) models \citep[e.g.][and references therein]{1980ApJ...239L..93D,1999APh....12..169B,2008MNRAS.385.1211P,2011A&A...527A..99E} for the origin of radio haloes \citep[e.g.][]{2011ApJ...728...53J,2017MNRAS.472.1506B}. On the other hand, in the case of radio relics, the non-detection of cosmic-ray protons directly challenges the shock acceleration mechanisms in the ICM \citep[e.g.][]{va14relics}. Here, we focus on the lack of shock-accelerated of cosmic-ray protons and, therefore, we point the reader to the references given above for further information on the other topics.

The collocation of shock waves, observed in X-rays, and radio relics \citep[e.g.][]{2013MNRAS.433.1701O,2013PASJ...65...16A,2016MNRAS.460L..84B,2016MNRAS.463.1534B,2018MNRAS.478.2218H} gave rise to the idea that ICM electrons are shock accelerated to high energies, where they emit synchrotron radiation \citep{ensslin1998}. The originally proposed acceleration mechanism is diffusive shock acceleration \citep[DSA, see][for physical details]{1978ApJ...221L..29B,1978MNRAS.182..147B,1978MNRAS.182..443B,1983RPPh...46..973D,1987PhR...154....1B,2014IJMPD..2330007B,bykov2019review,marcowith2020review}. While this picture is commonly accepted, some open questions remain. In several cases, the Mach number estimates from X-ray and radio observations yield different shock strengths \citep[][]{2014MNRAS.445.1213S,2015ApJ...812...49H,2016ApJ...818..204V,Hoang2017sausage}. In addition, while the radio power of some relics is explained by the shock acceleration from the thermal pool  \citep{locatelli2020dsa}, several other relics require shock re-acceleration \citep[e.g.][]{2014ApJ...785....1B,2017NatAs...1E...5V,Stuardi2019}. In this review, we concentrate on the \textit{missing $\gamma$-ray problem}, which we summarise in the following.

In principle, the protons in the ICM should be accelerated by the same shock waves that produce radio relics. Eventually, the accelerated cosmic-ray protons collide with the thermal protons of the ICM. These inelastic collisions would produce charged and neutral pions that further decay into electrons, positrons, neutrinos and $\gamma$-rays \citep{1999APh....12..169B}:
\begin{align}
 p+p &\rightarrow \pi^{+/-} + \pi^0 + \mathrm{anything} \label{eq::pp} \\
 \pi^{+/-} &\rightarrow \mu^{+/-} + \nu_{\mu} \label{eq::pisign} \\
  \mu^{+/-} &\rightarrow e^{+/-} + \bar{\nu}_{\mu} ({\nu}_{\mu}) + \nu_e (\nu_e) \label{eq::mu}  \\
 \pi^0 &\rightarrow 2 \gamma \label{eq::pizero} .
\end{align}
Hence, if the shock waves in the ICM are efficient proton accelerators, galaxy clusters should be observable in $\gamma$-rays\footnote{We note that this kind of $\gamma$-ray signal is also expected from radio halos, if they are produced by the so-called \textit{secondary models} \citep{1980ApJ...239L..93D}.}, (see Fig. \ref{fig::maps}). However, such signal has not been detected so far \citep[e.g.][]{ackermann14,ackermann15,ackermann16,Huber2013}. This non-detection poses a fundamental challenge on the shock acceleration mechanisms in the ICM. Several solutions have been proposed to explain the non-detection of $\gamma$-rays: modifying the cosmic-ray proton distribution in clusters \citep[e.g.][]{2011A&A...527A..99E,2013MNRAS.434.2209W,2014MNRAS.438..124Z,2017MNRAS.465.4800P}, revising the particle acceleration efficiencies \citep[e.g.][]{va14relics,2015MNRAS.451.2198V,ka12,2013MNRAS.435.1061P} and a better understanding of the microphysical processes of the shock acceleration mechanism \citep[e.g.][]{2014ApJ...783...91C,2014ApJ...794...46C,2014ApJ...794...47C,Guo_eta_al_2014_I,Guo_eta_al_2014_II,Ha2018protons,Ryu2019,2017JKAS...50...93K,2018ApJ...856...33K,2020JKAS...53...59K}. 

Reviewing all of these topics is beyond the scope of this work. Hence, we focus on the advances and implications of the latter, as new theoretical models on the microphysics at work during the shock acceleration are able to explain the lack of $\gamma$-rays. This paper is structured as follows: in Sec. \ref{sec::observations}, we give an overview of the observational limits obtained with modern $\gamma$-ray observatories. In Sec. \ref{sec::shocksandprotons}, we put forward the latest advances of theoretical and numerical modelling of proton shock acceleration and their effect on the associated $\gamma$-ray emission. Before the conclusion, we briefly give two alternative perspectives on the missing $\gamma$-ray problem in Sec. \ref{sec::additional}. In Sec. \ref{sec::summary}, we conclude the review and give an outlook on the future of $\gamma$-ray observations in the ICM.

\section{Observations of $\gamma$-rays in the ICM}\label{sec::observations}

In the search for $\gamma$-rays, both space-based and ground-based telescope are used \citep[e.g.][]{ackermann14,2014ApJ...795L..21G,zand14,aha09,alek10,alek12,alek14,ahnen16}. While ground-based observations mostly probe the energy range above $\gtrsim 1  \ \GeV$, space-based observations are mostly looking at energies between $\sim 100 \ \MeV $ and $\sim 1 \ \TeV$. Theoretical estimates yield that the $\gamma$-ray emission in the ICM is expected to be the strongest around $\sim 1 \ \GeV$ \citep{2017MNRAS.472.1506B,2010MNRAS.409..449P}, hence, favoring space-based observations to detect $\gamma$-rays. Though, ground- and space-based observations do not compete but very well complement each other \citep[e.g.][]{arl12}. However, none of them has ever reported the detection of a $\gamma$-ray signal from the ICM. Hence, they all provide upper limits on the $\gamma$-ray flux expected from cosmic-ray protons in the ICM. The deepest upper limits have been set by the Large Area Telescope\footnote{http://www-glast.stanford.edu/} on board of the Fermi Satellite\footnote{https://fermi.gsfc.nasa.gov/} (from here on Fermi-LAT). In the following, we will briefly review these limits and we point to the literature for other works. 

Since its launch in 2008, the Fermi-LAT has been surveying the $\gamma$-ray sky between $20 \ \MeV$ and $300 \ \GeV$ \citep[e.g.][for technical details]{2009ApJ...697.1071A,ack12tech} and, hence, it perfectly covers the desired energy range around $\sim 1 \ \GeV$ \citep[e.g.][]{2017MNRAS.472.1506B,2010MNRAS.409..449P}. Using Fermi-LAT's 4 year all-sky data, \cite{ackermann14} searched for spatially extended $\gamma$-ray emission in a sample of 50 X-ray bright galaxy clusters that included both cool-core and non-cool-core clusters. Using the cosmic-ray model by \cite{2010MNRAS.409..449P}, they thoroughly analysed the sample. Yet, they did not report any $\gamma$-ray emission attributed to the ICM. Hence, they presented upper limits for the $\gamma$-ray flux above $500 \ \MeV$ that are between $0.5$ and $22.2 \cdot 10^{-10} \ \mathrm{ph \ cm^{-2} \ s^{-1}}$. 
 
More extended searches targeted the $\gamma$-ray emission coming from individual clusters, namely the Virgo cluster \citep{ackermann15} and the Coma cluster \citep{ackermann16}. Albeit, these observations used a larger energy range, i.e. $\ge 100 \ \MeV$, and, in the case of Coma, 6 years of data, they did not report any $\gamma$-ray signal. Hence, the $\gamma$-ray flux limits are $5.2 \cdot 10^{-9} \ \mathrm{ph \ cm^{-2} \ s^{-1}}$ and $1.2 \cdot 10^{-8} \ \mathrm{ph \ cm^{-2} \ s^{-1}}$ for Coma and Virgo, respectively. In an independent work, \cite{zand14} used 63 months of Fermi-LAT observations to analyse the Coma cluster in the energy $100 \ \MeV$ to $100 \ \GeV$. For different physical scenarios, they found upper limits in the range of $10^{-10}$ to $10^{-9} \ \mathrm{ph \ cm^{-2} \ s^{-1}}$.
 
\cite{2013AA...560A..64H} used the stacking of $\gamma$-ray counts maps \citep[also see][]{2012A&A...547A.102H} to estimate statistical upper limits on a sample of 53 galaxy clusters taken from the HIFLUGS catalog \citep{Reiprich:2002}. For energies between $1$ and $300 \ \GeV$, they found an upper limit of a few $10^{-11} \ \mathrm{ph \ cm^{-2} \ s^{-1}}$ for their whole sample . Furthermore, they performed separate stacking analyses for cool-core and non-cool-core clusters, that yielded upper limits of $5.7$ and $2.7 \cdot 10^{-11} \mathrm{ph \ cm^{-2} \ s^{-1}}$, respectively. \cite{2014ApJ...795L..21G} performed an independent stacking analysis using 78 nearby clusters. In the $0.8$ to $100 \ \GeV$ energy band, they recorded an upper limit of $2.3 \cdot 10^{-11} \ \mathrm{ph \ cm^{-2} \ s^{-1}}$.
 
Overall, these non-detections of $\gamma$-rays produced by inelastic collisions between cosmic-ray protons and thermal protons in the ICM constrain the total content of cosmic-ray protons in the ICM. In general, the ratio between the cosmic-ray proton pressure and the total gas pressure is limited to below $1 \ \%$ \citep{2016MNRAS.459...70V}.

\section{Advances from theory and simulations} \label{sec::shocksandprotons}

A challenge for theoretical and numerical works is to explain the non-detection of $\gamma$-rays. In a different perspective, the above quoted limits on the $\gamma$-ray flux provide an ideal testcase for shock acceleration models in the ICM. The major challenge of any new model is to explain the non-detection cosmic-ray protons while still accelerating enough cosmic-ray electrons to produce visible radio relics. 

\begin{figure}[h]
 \includegraphics[width = \textwidth]{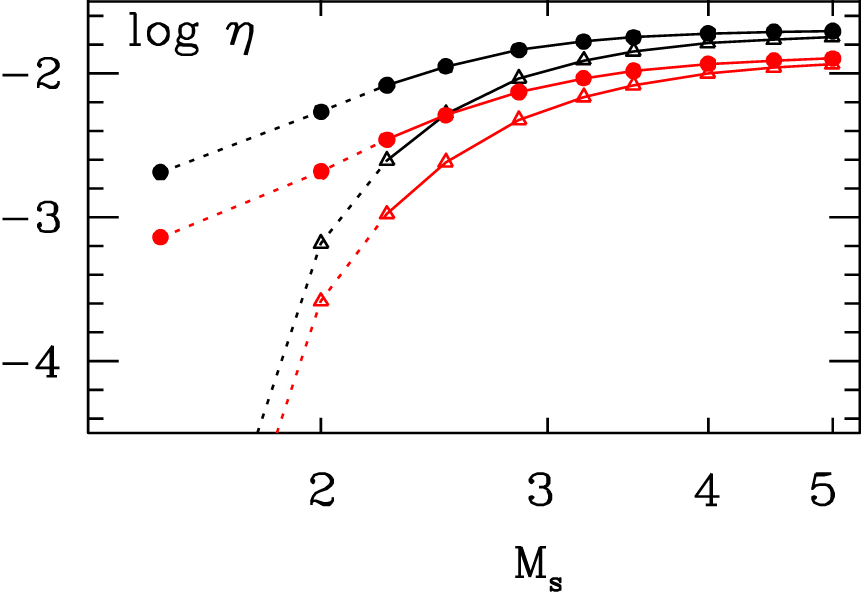}
 \caption{Results given in \cite{Ryu2019}: cosmic-ray acceleration efficiencies as a function of Mach number. The red and black lines give the efficiencies for injection moments of $p_{\mathrm{inj}} = 3.5 \cdot p_{\mathrm{th}}$ (red) and $p_{\mathrm{inj}} = 3.3 \cdot p_{\mathrm{th}}$ (red), where $p_{\mathrm{th}}$ is the momentum of the thermal post-shock electrons. The circles show the results, if the injection momentum is assumed as a lower bound for computing the injection fraction, while the triangles assume the threshold energy for the pion-production reaction, $780 \ \MeV/c$, as a lower bound. The points at $M_\mathrm{s} = 2$ and $M_{s} = 1.5$ are not able to accelerate protons. However, they are plotted for completeness.}
 \label{fig::ryu2019}
\end{figure}

The $\gamma$-ray signal strongly depends on the amount of energy injected into the cosmic-ray protons by the shock wave. Hence, by reducing the injected energy, it is possible to decrease the corresponding $\gamma$-ray signal. One possibility to do this is to lower the shock acceleration efficiencies, $\eta$. For shock with Mach number $M$, the acceleration efficiency is defined as the ratio of the injected cosmic-ray energy flux, $f_{\mathrm{CR}}$, and the total kinetic energy flux dissipated by the shock, $f_{\mathrm{tot}}$, \citep[e.g.][]{ryu2003}:
\begin{align}
 \eta(M) = \frac{f_{\mathrm{CR}}(M)}{f_{\mathrm{tot}}(M)}.
\end{align}

Using cosmological simulations, \cite{2016MNRAS.459...70V} tested a variety of acceleration efficiencies (that we discuss further below). They found that the efficiencies, available at that time, were too large and produced visible $\gamma$-rays. Furthermore, they showed that an overall constant acceleration efficiency of $\eta = 10^{-3}$ would be required to make galaxy clusters undetectable to current instruments. A word of caution: simply lowering the acceleration efficiencies is difficult to be brought in line with the observations of radio relics. Especially as low acceleration efficiencies already have difficulties in explaining the radio power of several observed relics \citep[e.g.][]{2014ApJ...785....1B,2017NatAs...1E...5V,Stuardi2019}.

\begin{figure}[h]
 \includegraphics[width = \textwidth]{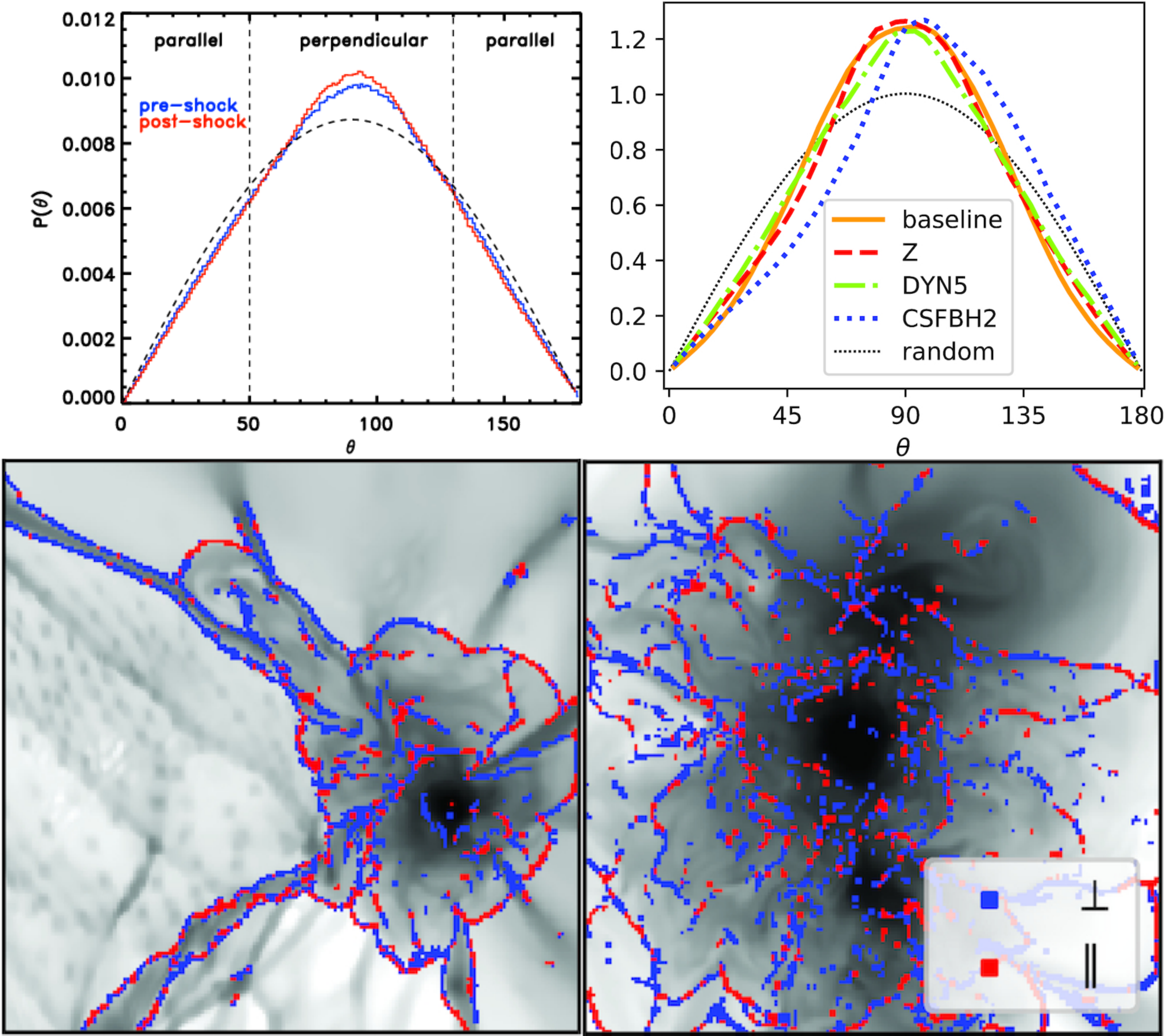}
 \caption{Results given in \cite{2017MNRAS.464.4448W} and \cite{Banfi2020}. Top left: pre-(blue) and post-shock (red) obliquity distribution of the relic simulated in \cite{2017MNRAS.464.4448W} compared to the distribution of angles for a random distribution. Top right: obliquity distribution measured for different magnetic field seeding models \citep[colored lines, see][for details]{Banfi2020} compared to the distribution of angles between two random vectors in a three dimensional space (black dotted lines). Bottom panel: spacial distribution of obliquities in two galaxy clusters. The gray color gives the ICM density. The blue squares mark quasi-perpendicular shocks, while the red squares highlight quasi-parallel shocks.}
 \label{fig::obli}
\end{figure}

Pioneering works by \cite{kj07} used 1D diffusion convection equations of shocks to estimate the shock acceleration efficiencies. Using a similar model, \cite{2007A&A...473...41E} derived acceleration efficiencies for different post-shock temperatures. However, both of these are above the $\eta = 10^{-3}$ limit given by \cite{2016MNRAS.459...70V}. By including more microphysical effects, \cite{2013ApJ...764...95K,2014ApJ...783...91C,Ryu2019} updated the efficiencies given in \cite{kj07}, always yielding lower efficiencies. \cite{2013ApJ...764...95K} included energy dissipation due to Alfven waves that are amplified at the shock. \cite{2014ApJ...783...91C} took into account the effect of the shock obliquity on the acceleration process (see discussion below), which gives efficiencies that are half the efficiencies derived by \cite{2013ApJ...764...95K}. Finally, \cite{Ryu2019} also included the dynamical feedback of cosmic-ray pressure on the shock. This model yields acceleration efficiencies, see Fig. \ref{fig::ryu2019}, that are lower than the ones of its predecessors and that fall below the $\eta = 10^{-3}$ limit given by \cite{2016MNRAS.459...70V}. Yet, it was shown that these efficiency reduce the $\gamma$-ray emission significantly, but not enough to make galaxy clusters invisible in $\gamma$-rays \citep[e.g.][and below]{wittor2020gammas}.

In the recent years, several kinetic simulations found that not all type of shocks are able to efficiently accelerate cosmic-ray protons \citep{2014ApJ...783...91C,Ha2018protons}. Shocks can be characterized by two properties: their strength, given by the Mach number, and their orientation to the underlying magnetic field, given by the shock obliquity. The latter is defined as the angle between shock normal, $\mathbf{n_{shock}}$, and magnetic field, $\mathbf{B}$:
\begin{align}
 \theta = \arccos\left( \frac{ \mathbf{n_{shock}} \cdot \mathbf{B} }{ \left| \mathbf{n_{shock}} \right| \left|\mathbf{B} \right|  }  \right)
\end{align}

Using the obliquity, shocks are classified as either quasi-parallel, i.e. $\theta \lesssim 45^{\circ}$, or quasi-perpendicular, i.e. $\theta \gtrsim 45^{\circ}$. The Mach number states if a shock is supercritical or not \citep[e.g.][]{1955RSPSA.233..367M}. Following the definition given in \cite{Ha2018protons}, a shock is supercritical if its Mach number is above $ \gtrsim 2.25$.

\cite{2014ApJ...783...91C} showed that the cosmic-ray proton injection is only efficient in quasi-parallel shocks.  In the picture of DSA, particles gain energy while they are scattered of magnetic inhomogeneities back and forth across the shock front. Hence, one of the keys for efficient DSA are the magnetic inhomogeneities in the shock upstream and downstream. However, only at quasi-parallel shocks, cosmic-ray protons are able to induce magneto-hydrodynamical instabilities that again cause such inhomogeneities to grow. At quasi-perpendicular shocks, the growth of such instabilities  is prevented for two reasons \citep[also see the discussion in][]{2014ApJ...783...91C}. Most important, protons move only for one gyroradius into the upstream and, hence, the time available for any cosmic-ray driven perturbation to grow is significantly reduced. Secondly, the particle anisotropies are mostly along the flow, which prohibits the growth of most instabilities. However, quasi-perpendicular shocks might still be able to re-accelerate an existing cosmic-ray proton population by DSA.

Several works measured the distribution of shock obliquities in cosmological simulations \citep{2016Galax...4...71W,2017MNRAS.464.4448W,Roh2019,Banfi2020}. These works showed that, to first order, the distribution of shock obliquities in the ICM follows the distribution of angles between two random vectors in a three dimensional space, that is $\propto \sin \left( \theta \right)$, see in Fig. \ref{fig::obli}. Hence, only $\sim 33 \ \%$ of all shocks are expected to be quasi-parallel. Additionally, the excess of quasi-perpendicular shocks over quasi-parallel shocks is somewhat larger than for a pure random distribution, as shock waves amplify the perpendicular component of the magnetic field leading to more quasi-perpendicular shocks \citep{2017MNRAS.464.4448W}. Using a larger set of cosmological simulations, \cite{Banfi2020} measured the obliquity distribution in clusters, filaments and voids. They found that locally this excess can be up to factors of $\sim 5$ larger than for the random case. Furthermore, the  excess of quasi-perpendicular shocks is maximized for Mach numbers with $5 \le M \le 20$ \citep[also see][]{2016Galax...4...71W} but it typically depends on the pre-shock temperature. Furthermore, the excess is maximized around filaments, where the shock is perpendicular to the filaments, see Fig. \ref{fig::obli}. Here, the magnetic field aligns with the filaments because of an equilibrium process that aligns local magnetic field with the density gradient as found by \cite{soler2017}.

However inside clusters, turbulent motions randomize the obliquity distribution again. Hence, to first order, about $\sim 33 \ \%$ of all shocks are quasi-parallel. Restricting the cosmic-ray proton acceleration to quasi-parallel shocks reduces the associated $\gamma$-ray emission by factor of about $\sim 3$. Yet, this drop is not enough to explain the non-detection of $\gamma$-rays \citep{2016Galax...4...71W,2017MNRAS.464.4448W,wittor2020gammas}. It is worth to notice, that the obliquity requirement does not affect the emission from radio relics. In a complementary work, \cite{Guo_eta_al_2014_I} found that cosmic-ray electrons are only efficiently accelerated by quasi-perpendicular shocks. Therefore, most shocks in the ICM, i.e. $\sim 66 \ \%$, are able to accelerate cosmic-ray electrons which leaves the power of relics almost unaltered \citep{2016Galax...4...71W,2017MNRAS.464.4448W}. 

\begin{figure}[h]
 \includegraphics[width = \textwidth]{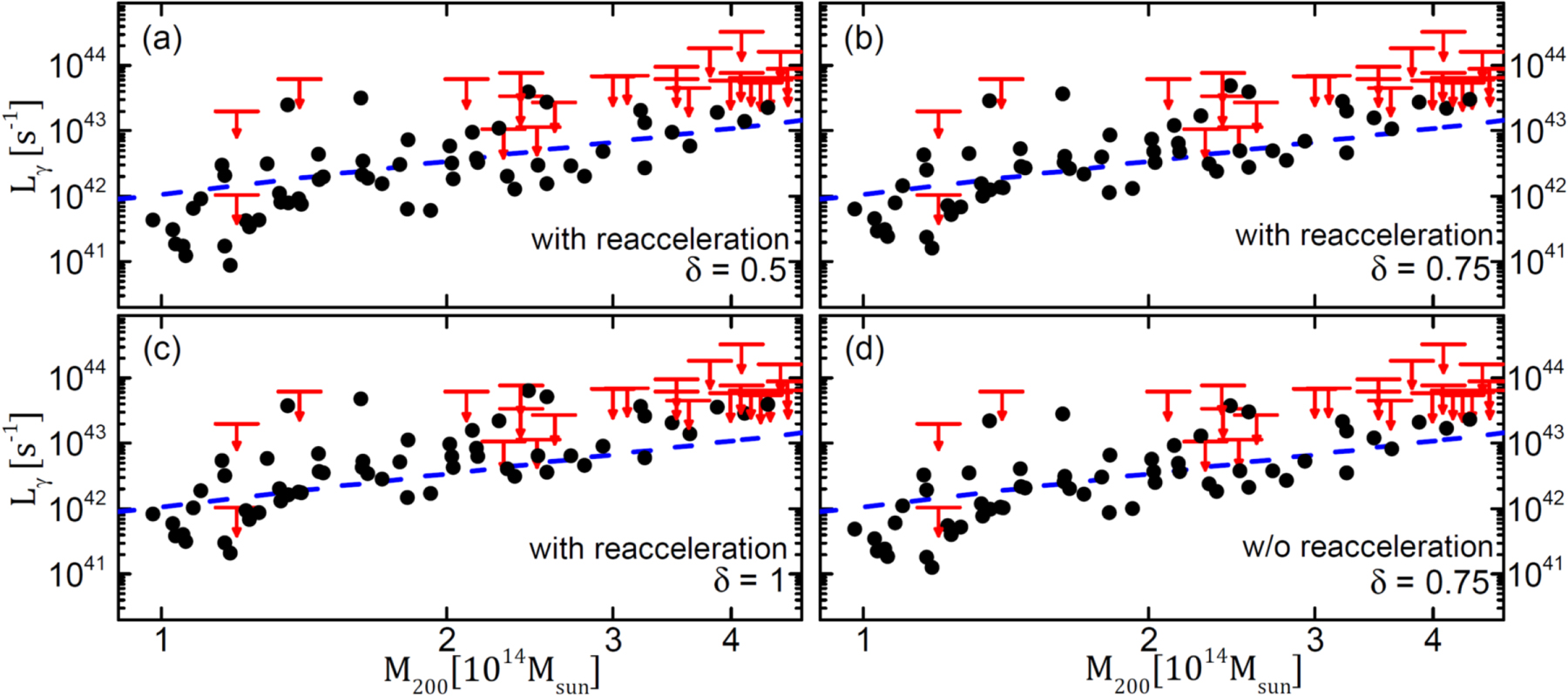}
 \caption{Results given in \cite{Ha2019gammas}: The various panels compare the $\gamma$-ray luminosities, as a function of cluster mass, of the simulated sample (black dots) with the upper limits produces by the Fermi-LAT \citep[red crosses][]{ackermann14}. The luminosities have been computed/measured in the energy band $[0.5,200] \ \GeV$. The different panels give the results for different spatial distributions of the cosmic-ray protons, i.e. $n_{\mathrm{CR}}(r) \propto n_{\mathrm{gas}}(r)^{\delta}$. The first three panels, (a)-(c), show the luminosities with re-acceleration included, while in the last panel, (d), re-acceleration was neglected. The blues lines give the mass-luminosity relation, $L_{\gamma} \propto M_{200}^{5/3}$. }
 \label{fig::ha2019}
\end{figure}

Using PIC simulations, \cite{Ha2018protons} studied on the proton acceleration by quasi-parallel low Mach number shocks in high $\beta$ plasmas. In their set-up, they assumed a shock obliquity of $\theta = 13^{\circ}$ and a proton to electron mass ration of 100. They showed that under such conditions only supercritical shocks are able to accelerate cosmic-ray protons, i.e. the acceleration is quenched for shocks with Mach numbers below $M \lesssim 2.25$. However, the critical Mach number depends on the upstream plasma parameters, such as the obliquity or the plasma $\beta$ \citep[e.g.][]{1984JPlPh..32..429E}. Furthermore, hybrid simulations of weak shocks in high $\beta$ plasmas indicated that the ion distribution and, hence, the critical Mach number are sensitive to the cosmic helium abundance \citep{bykov2019review}. Yet, constraining the exact helium abundance remains as a task for the next generation of X-ray observatories \citep[e.g.][]{2018SPIE10699E..1GB}. Hence, despite the importance of the measurement by \cite{Ha2018protons}, the exact value of the critical Mach number might differ, if more realistic values for the proton to electron mass ration and the helium abundance are taken into account.

Most shocks in the ICM are weak \citep{ryu2003,Vazza:2009_shock,2013ApJ...765...21S,Ha2018mergershocks,wittor2019pol} and, hence, the supercritical-requirement reduces the number of shocks, that are able to accelerate protons, significantly. However, \cite{wittor2020gammas} found that the supercritical criteria alone does not reduce the $\gamma$-ray emission enough to become invisible. Despite the small number of supercritical shocks, they argue that most of the cosmic-ray energy flux is processed by these few supercritical shocks, on average $\sim 53 \ \%$, and, hence, the associated $\gamma$-ray emission is only reduced by a factor of about $\sim 2$. This reduction is not enough to explain the non-detection of $\gamma$-rays.

In a pioneering work, \cite{Ha2019gammas} tested if shock accelerated cosmic-ray protons become invisible to the Fermi-LAT if: a) the shock efficiencies are the ones derived by \cite{Ryu2019}, and b) only supercritical and quasi-parallel shocks are able to accelerated cosmic-ray protons. Using a particle-mesh/Eulerian cosmological hydrodynamic \citep{ryu1993}, they simulated 58 galaxy clusters that cover a mass range of $10^{14}-5\cdot 10^{14} \ \Msun$. In their simulation, they evolved the magnetic fields passively using the Biermann Battery mechanism \citep{Biermann}. As they did not self-consistently follow the transport of cosmic-ray protons, they computed the $\gamma$-ray emission in post-processing. Therefore, they identified all shocks in their simulation at redshift $z = 0$ and measured their properties such as strength, obliquity and energy flux. By assuming a radial distribution of cosmic-ray protons, they computed the volume-integrated momentum distribution of cosmic-ray protons that is produced by the detected shocks. Finally, they estimated the associated $\gamma$-ray emission by applying the formalisms of \cite{pe04,kelner2006}. They find that only $\sim 7 \ \%$ of the kinetic energy flux is processed by supercritical and quasi-parallel shocks. In addition, the shocks that are able to accelerate cosmic-ray protons have an average strength of $M \approx 2.8 - 3.3$. The $\gamma$-ray emission in the simulation was compared to the upper limits of the cluster sample given in \cite{ackermann14}, see Fig. \ref{fig::ha2019}. Indeed, the $\gamma$-ray emission of the simulated clusters falls below the upper limits given by the Fermi-LAT. 

\begin{figure}[h]
 \includegraphics[width = \textwidth]{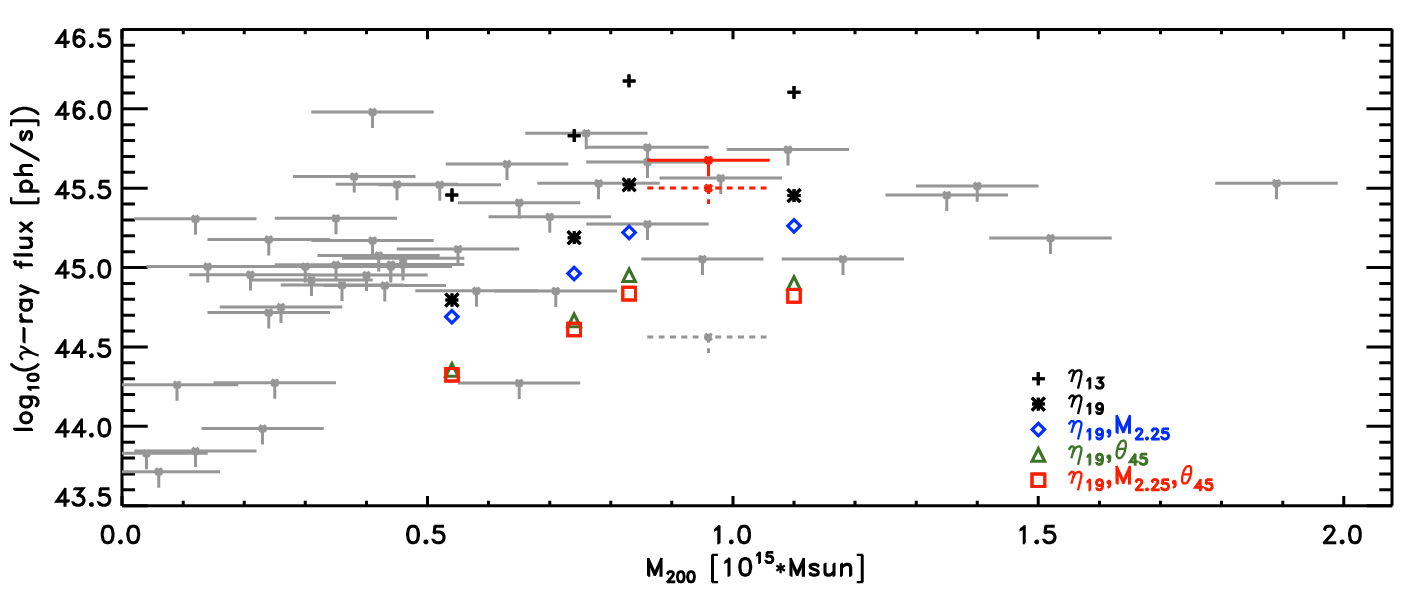}
 \caption{Results given in \cite{wittor2020gammas}: The plot compares the total $\gamma$-ray flux the simulated sample of \cite{wittor2020gammas} with the upper limits given by the Fermi-LAT. Specifically, the grey crosses give the limites in the energy range $[0.5,200] \ \GeV$ \cite{ackermann14}. The red solid cross marks the deeper upper limit of the COMA cluster \cite{ackermann16}, while the red dashed cross is the expected upper limit after ten more years of Fermi-LAT observations of COMA. The different colored symbols give the $\gamma$-ray fluxes computed for the simulated sample. The model that combines the acceleration efficiencies from \cite{Ryu2019} with acceleration by supercritical and quasi-parallel shocks only is given by the red squares. The other models are: all shocks (black asterisks), only supercritical shocks (blue diamonds) or only quasi-parallel shocks (green triangles) are able to accelerate cosmic-ray protons. The black crosses display the $\gamma$-ray flux obtained for the ``old'' acceleration efficiencies calculated by \cite{2013ApJ...764...95K}. (The plot has been modified from its original version to fit the style of this paper).}
 \label{fig::gamma_vs_limits_all_cluster}
 \end{figure}

In a complementary work, \cite{wittor2020gammas} combined cosmological simulations and La\-grang\-ian tracer particles to study the cosmic-ray proton acceleration in four massive galaxy clusters, i.e. $5.4 \cdot 10^{14}- 1.1 \cdot 10^{15} \ \Msun$ (see Fig. \ref{fig::maps}). In their modelling, they applied the acceleration efficiencies derived by \cite{Ryu2019}. In addition, they measured the reduction of the $\gamma$-ray flux if a) only supercritical shocks, b) only quasi-parallel shocks and c) only both supercritical and quasi-parallel shocks are able to accelerate cosmic-ray protons. In agreement with the results of \cite{Ha2019gammas}, they found that, using the acceleration efficiencies from \cite{Ryu2019}, the associated $\gamma$-ray emissions is invisible to the Fermi-LAT, if only supercritical and quasi-parallel shocks are able to accelerate cosmic-ray protons, see Fig. \ref{fig::gamma_vs_limits_all_cluster}. Furthermore, they found that most of the energy flux is processed by strong shocks and, hence, the key ingredient, to explain the non-detection of $\gamma$-rays, is the obliquity cut.

In principle, shocks can also re-accelerate cosmic-ray protons. For weak shocks, the shock acceleration efficiencies are larger \citep[e.g.][]{2013ApJ...764...95K} and, hence, they can inject a larger fraction of cosmic-rays. Furthermore, it is still uncertain whether re-acceleration works at every type of shock or if it is restricted to supercritical \citep{Ha2018protons} and quasi-parallel \citep{2014ApJ...783...91C}, as well. Hence, re-accelerated cosmic-ray protons might produce a $\gamma$-ray signal that is above the Fermi-limits. 

\cite{Ha2019gammas} also tested their model for cosmic-ray re-acceleration. In their model, cosmic-protons are shock accelerated at three fixed periods. While the first acceleration event is pure thermal acceleration, the second and third event also include re-acceleration. Their estimation yields that about $\sim 6 - 8 \ \%$ of pre-existing cosmic-ray protons undergo re-acceleration. Assuming that only quasi-parallel and supercritical shocks are able to re-accelerate cosmic-ray protons, they found that the energy stored in cosmic-ray protons increases about $\sim 60 \ \%$ on average. Yet, the associated $\gamma$-ray emission still remains below the Fermi-limits, see Fig. \ref{fig::ha2019}. For completeness, they tested the effect of re-accelerating when neglecting the supercritical or obliquity criteria. Their estimates showed that, when relaxing the supercritical criteria, the total cosmic-ray proton energy grows by $\sim 90 \ \%$ on average. On the other hand, if the re-acceleration is independent of the shock obliquity, the cosmic-ray proton energy increases significantly and becomes too large to be compatible with upper limits set by the Fermi-LAT.

\section{A different perspective}\label{sec::additional}

Before concluding this work, we want to shortly highlight two different perspectives on the missing $\gamma$-ray problem. First, we present an other possibility to reduce the amount of cosmic-ray protons inside clusters and, hence, to lower the associated $\gamma$-ray signal. Second, we want to briefly point towards neutrinos which are an other tracer of cosmic-ray protons in the ICM.

As the $\gamma$-ray signal is produced by collisions between cosmic-ray protons and thermal protons, it strongly depends on the amount of cosmic-ray protons inside the cluster volume. More specifically, it is proportional to the number density squared, $\propto n^2$. Hence, if the amount of cosmic-ray protons living in the clusters' central regions is reduced, the associated $\gamma$-ray signal drops significantly. \cite{2011A&A...527A..99E} were the first to propose that cosmic-ray protons could stream out of the central regions and, hence, flattening the cosmic-ray protons profiles. Though, it is difficult to explain the lack of $\gamma$-ray emission in galaxy clusters that host radio halos with cosmic-ray streaming \citep[e.g.][and references therein]{vanweeren2019review}. However, cosmic-ray streaming is complex topic as it depends on the interactions between cosmic-rays and plasma waves and turbulence. Hence, we point the interested reader to sophisticated works and the references therein: \cite{2017MNRAS.465.4800P,2011A&A...527A..99E,2013MNRAS.434.2209W,2014MNRAS.438..124Z,2018MNRAS.473.3095W}. 

The pions, that are produced by the inelastic collisions of cosmic-ray protons and protons, do not only decay into $\gamma$-rays but also into neutrinos (see Eq. \ref{eq::pisign} and \ref{eq::mu}). Hence, these neutrinos provide an other observable that could reveal the presence of cosmic-ray protons in the ICM. Several works have focused on estimating the neutrino flux coming from galaxy clusters \citep[e.g.][]{Murase2008,Murase2013,MuraseWaxman2016,Zandanel2015}. However, these works also showed that shock accelerated cosmic-ray protons do not produce significant fraction of the neutrino flux measured by IceCube\footnote{https://icecube.wisc.edu/}. Nevertheless, \cite{Ha2019gammas} computed the neutrino flux in their sample of simulated galaxy clusters (see Sec. \ref{sec::shocksandprotons}). They estimated that the neutrino fluxes in their sample are significantly below both the atmospheric neutrino fluxes at $E \le 1 \ \mathrm{TeV}$ and the IceCube flux at $E = 1 \ \mathrm{PeV}$ \cite[][respectively]{Richard2016,Aartsen2014}. Hence, they concluded that the detection of neutrinos, coming from the collisions of cosmic-ray protons and protons, is more than unlikely with the current and future generation of ground based neutrino telescopes, i.e. IceCube \citep{IceCubeRevPaper2010}, Super-Kamiokande \citep{2019ApJ...887L...6H} or future Hyper-Kamiokande \citep{2011arXiv1109.3262A}.

\section{Will there be a detection in the future?}\label{sec::summary}

In this review, we have reviewed the latest advances in understanding the \textit{missing $\gamma$-ray problem}, which fundamentally challenges the shock acceleration mechanisms at work in the ICM. Thanks to a collaborative effort of theoretical predictions, numerical simulations as well as space-based and ground-based observations, it is now possible to explain the non-detection of $\gamma$-rays associated to shock accelerated cosmic-ray protons: Recent studies showed that if both the energy dissipation due to Alfven waves and the dynamical feedback of the cosmic-ray pressure on the shock are included to derive shock acceleration efficiencies \citep{Ryu2019}, then they efficiencies become significantly smaller than previously estimated. If these new efficiencies are paired with the fact that only supercritical \citep{Ha2018protons} and quasi-parallel \citep{2014ApJ...783...91C} shocks are able accelerated cosmic-ray protons, then the expected $\gamma$-ray signal from the ICM drops below the upper limits given by the Fermi-LAT \citep{Ha2019gammas,wittor2020gammas}. 

Yet, only a direct detection of a $\gamma$-ray signal associated with cosmic-ray protons in the ICM will allow to pin-point the exact shock acceleration mechanism. Though, estimates by \cite{wittor2020gammas} showed that, even after ten more years of operation, the Fermi-LAT will most-likely not detect such a signal, see Fig. \ref{fig::gamma_vs_limits_all_cluster}. Hence, all hopes to detect $\gamma$-rays in the future lie with the next generation of $\gamma$-ray telescopes. 

The Cherenkov Telescope Array\footnote{https://www.cta-observatory.org/} (CTA) is the next ground-based $\gamma$-ray that is currently being build in Paranal, Chile, and on La Palma, Spain, and its construction is planned to be completed in 2025. Yet, in the desired energy range, i.e.$10^2-10^4 \ \MeV$, the CTA is expected to be less sensitive than the Fermi-LAT \citep{CTA}. Therefore, it will most-likely not be able to detect the missing $\gamma$-rays.

However, the proposed space mission: All-sky Medium Energy Gamma-ray Observatory\footnote{https://asd.gsfc.nasa.gov/amego/} \citep[AMEGO][]{amego} will have the capabilities to detect the missing $\gamma$-rays. If accepted, AMEGO will observe in the energy range $0.2 \ \MeV$ to $10 \ \GeV$ and it sensitivity will be factors of $\sim 10 - 30$ below Fermi-LAT's sensitivity. This could potentially be enough to finally detect the missing $\gamma$-rays. Yet, also a non-detection with AMEGO would provide relevant information about the cosmic-ray proton spectrum. Some models of cosmic-ray proton re-acceleration by weak shock ensembles with long-wavelength magneto-hydrodynamical waves produce a soft asymptotic cosmic-ray proton spectrum \citep[e.g.][]{bykov2019review}. Such a soft spectrum can help to understand the non-detection by the Fermi-LAT, yet it will be difficult to be detected even with AMEGO.

\section*{Acknowledgements}\label{sec::acknowledgements}

The author would like to thank referee for the helpful comments, which
helped to improve the quality of this review.
 
Furthermore, the author would like to thank D. Ryu, J. Ha and S. Banfi for providing the figures associated to their work. The author would like to thank F. Vazza and M. Br\"uggen for introducing him to the subject of proton shock acceleration the ICM and for fruitful discussions about the topic. 
 
D.W. is funded by the Deutsche Forschungsgemeinschaft (DFG, German Research Foundation) - 441694982. The author gratefully acknowledges the Gauss Centre for Supercomputing e.V. (www.gauss-centre.eu) for supporting this project by providing computing time through the John von Neumann Institute for Computing (NIC) on the GCS Supercomputer JUWELS at J\"ulich Supercomputing Centre (JSC), under project no. hhh44.

\section*{References}

\bibliography{mybib}

\end{document}